\documentclass[final,3p,times]{elsarticle}

\usepackage{amssymb}

\usepackage{subfigure}
\usepackage[ansinew]{inputenc}
\usepackage{listings}
\usepackage{algorithmic}

\journal{Journal of Systems and Software}

\begin{document}

\begin{frontmatter}

\title{A methodology to automatically optimize dynamic memory managers applying grammatical evolution}

\author[addr1]{Jos\'{e} L. Risco-Mart\'{i}n}
\author[addr2]{J. Manuel Colmenar}
\author[addr1]{J. Ignacio Hidalgo}
\author[addr1]{Juan Lanchares}
\author[addr3]{Josefa D\'{i}az}
\address[addr1]{Department of Computer Architecture and Automation, Universidad Complutense de Madrid, 28040 Madrid, Spain}
\address[addr2]{C. E. S. Felipe II, Universidad Complutense de Madrid, 28300 Aranjuez, Spain}
\address[addr3]{C.U. M\'{e}rida, Universidad de Extremadura, 06800 M\'{e}rida, Spain}

\begin{abstract}
Modern consumer devices must execute multimedia applications that exhibit high resource utilization. In order to efficiently execute these applications, the dynamic memory subsystem needs to be optimized. This complex task can be tackled in two complementary ways: optimizing the application source code or designing custom dynamic memory management mechanisms. Currently, the first approach has been well established, and several automatic methodologies have been proposed. Regarding the second approach, software engineers often write custom dynamic memory managers from scratch, which is a difficult and error-prone work. This paper presents a novel way to automatically generate custom dynamic memory managers optimizing both performance and memory usage of the target application. The design space is pruned using grammatical evolution converging to the best dynamic memory manager implementation for the target application. Our methodology achieves important improvements (62.55\% and 30.62\% better on average in performance and memory usage, respectively) when its results are compared to five different general-purpose dynamic memory managers.
\end{abstract}


\end{frontmatter}

\section{Introduction}
\label{sec:Introduction}

Nowadays, multimedia applications are mostly developed using C++. This kind of software programs tend to make intensive use of dynamic memory due to their inherent data management. However, in C++, dynamic memory is allocated via the operator \texttt{new()} and deallocated by the operator \texttt{delete()}, which are mapped directly to the \texttt{malloc()} and \texttt{free()} functions of the standard C library in most compilers. Therefore, the creation and destruction of objects is managed by a general-purpose memory allocator, which may provide good runtime and low memory usage for a wide range of applications \cite{Johnstone1999, Lea}.

However, using specialized \emph{Dynamic Memory Managers (DMMs)} that take advantage of application-specific behavior can dramatically improve application performance \cite{Barrett1993, Grunwald1993}. In this regard, three out of the twelve integer benchmarks included in SPEC (\texttt{parser}, \texttt{gcc}, and \texttt{vpr} \cite{SPEC2010}) and several server applications, use one or more custom DMMs \cite{Berger2001}.

On the one hand, studies have shown that dynamic memory management can consume up to 38\% of the execution time in C++ applications \cite{Calder1995}. Thus, the performance of dynamic memory management can have a substantial effect on the overall performance of C++ applications. On the other hand, new multimedia devices must rely on dynamic memory for a very significant part of their functionality due to the inherent unpredictability of the input data. These devices also integrate multiple services such as multimedia and wireless network communications which also compete for memory space. Then, the dynamic memory management influences the global memory usage of the system \cite{Atienza2006a}. Finally, energy consumption has become a real issue in overall system design due to circuit reliability and packaging costs \cite{Vijaykrishnan2003}. However, it has been recently proved that the DMM consumes only a 1\% of the total enery consumption by the memory subsystem usually in the execution of a given application \cite{Diaz2011}. Thus, the energy consumption by the DMM is not relevant on this case and the optimization of the dynamic memory subsystem has two goals that cannot be seen independently: performance and memory usage. There cannot exist a memory allocator that delivers the best performance and least memory usage for all programs. However, a custom memory allocator that works best for a particular program can be developed using grammatical evolution \cite{RiscoMartin2009b}.

To reach higher performance, programmers often write their own \textit{ad hoc} custom memory allocators as macros or monolithic functions in order to avoid function call overhead. This approach, implemented to improve application performance, is enshrined in the best practices of skilled computer programmers \cite{Meyers1995}. Nonetheless, this kind of code is brittle and hard to maintain or reuse, and as the application evolves, it can be difficult to adapt the memory allocator as the application requirements vary. Moreover, writing these memory allocators is both error-prone and difficult. Indeed custom and efficient memory allocators are complicated pieces of software that require a substantial engineering effort.

\begin{figure*}[ht]
\centering
\subfigure[Profiling of the application]{
\includegraphics[scale=0.618]{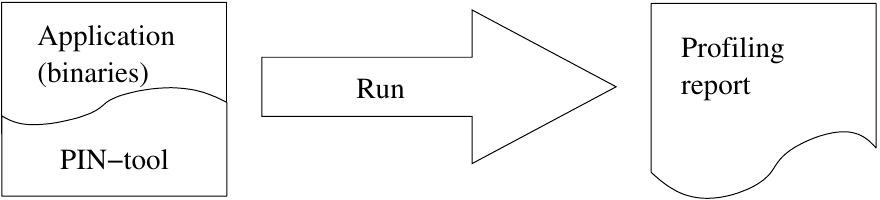}
\label{fig:Framework1}
}
\subfigure[Grammar generation]{
\includegraphics[scale=0.618]{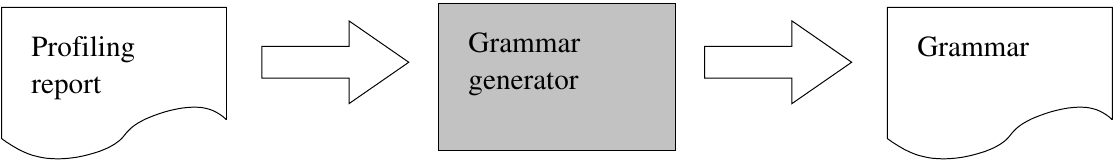}
\label{fig:Framework2}
}
\subfigure[Optimization]{
\includegraphics[scale=0.618]{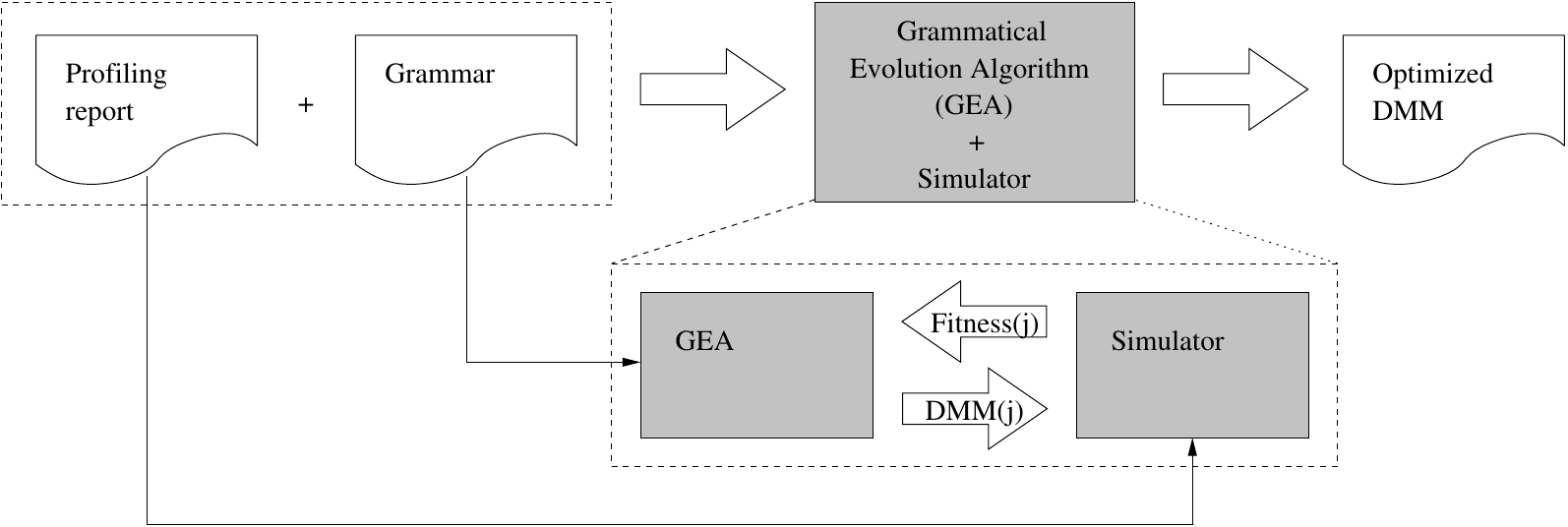}
\label{fig:Framework3}
}
\label{fig:Framework}
\caption{DMMs optimization flow. In the first phase, we generate an initial profiling of the de/allocation pattern. In the second phase, we automatically analyze the profiling report to generate the final grammar. Finally, in the third phase an exploration of the design space of DMMs implementation is performed using GE.}
\end{figure*}

In this work, we have developed a framework based on grammatical evolution to automatically design optimized DMMs for a target application, minimizing memory usage and maximizing performance. \figurename~\ref{fig:Framework} depicts the optimization process. First, as \figurename~\ref{fig:Framework1} shows, we run the application under study  together with an instrumentation tool, which logs all the required information into an external file: identification of the object created/deleted, operation (allocation or deallocation) object size in bytes and memory address. Since all the DMM exploration process is performed simulating the generated DMMs with the profiling report, this task must be done just once. In the following phase, as \figurename~\ref{fig:Framework2} shows, we automatically examine all the information contained in the profiling report, obtaining a specialized grammar for the target system. As a result, some incomplete rules in the original grammar (see Section \ref{sec:TheGrammar}), such as the different block sizes, are automatically defined according to the obtained profiling. To this end, we have developed a tool called \emph{Grammar Generator}. The last phase is the optimization process. As \figurename~\ref{fig:Framework3} depicts, this phase consists of a \textit{Grammatical Evolution Algorithm (GEA)} that takes the grammar generated in the previous phase and the profiling report of the application as inputs. GEA is supported by a DMM simulator that tests the behavior of every DMM generated by the grammar applied to the application. Our GEA is constantly generating different DMM implementations from the grammar file. When a DMM is generated ($\mathit{DMM(j)}$ in \figurename~\ref{fig:Framework3}), it is received by the DMM simulator. Then, the simulator emulates the behavior of the application, debugging every line in the profiling report. Such emulation does not de/allocate memory from the computer like the real application, but maintains useful information about how the structure of the selected DMM evolves in time. After the profiling report has been simulated, the DMM simulator returns back the fitness of the current DMM to the GEA. The fitness is computed as a weighted sum of the performance and memory usage by the proposed DMM for the target device and application under study. Finally, the DMM with lowest fitness is returned as solution (optimized DMM). 

The rest of the paper is organized as follows. First, Section \ref{sec:RelatedWork} describes some recent advances in the area of DMMs. Next, Section \ref{sec:DynamicMemoryManagement} defines the design space of memory allocators. Then, Section \ref{sec:TheSimulator} details the design and implementation of the DMM simulator, as well as some configuration examples. Section \ref{sec:TheGrammar} details how grammatical evolution is applied to the DMM optimization. Section \ref{sec:SetUp} shows our experimental methodology, presenting the six benchmarks selected, whereas Section \ref{sec:Experiments} shows the results for these benchmarks. Finally, Section \ref{sec:Conclusions} draws conclusions and future work.


\section{Related work}\label{sec:RelatedWork}


Several approaches have been presented in the last decade to design flexible and efficient infrastructures for building custom and general-purpose memory allocators \cite{Berger2001, Atienza2006a, Atienza2006}. All the proposed methodologies are based on high-level programming where C++ templates and object-oriented programming techniques are used. They allow the software engineer to compose both general-purpose and custom memory allocator mechanisms.
The aforementioned methodologies enable the implementation of custom DMMs from their basic parts (e.g., de/allocation strategies, order within pools, splitting, coalescing, etc.). In addition, \cite{Atienza2006a} and \cite{Atienza2006} provide a way to evaluate the memory usage and energy consumption, but at system-level. However, all the previously mentioned approaches require the execution of the target application to evaluate every candidate custom DMM, which is a very time-consuming task, especially if the target application requires human inputs (like video games). In this regard, \cite{Lo2004} and \cite{Teng2008} presented two DMM design frameworks that allow the definition of multiple memory regions with different disciplines. However, these approaches are limited to a small set of user-defined functions for memory de/allocation. Furthermore, the selection of the ``best'' DMM is based on a set of predefined rules and mono-objective search, respectively. Thus, new multi-objective approaches to measure performance and memory usage are needed when designing a custom or general-purpose DMM.

An initial design of the GEA proposed in this work was presented in \cite{RiscoMartin2009b} and the DMM simulator was described in \cite{RiscoMartin2010a} and \cite{RiscoMartin2011a}. However, the design space was defined in terms of previous DMM classifications performed by Berger \cite{Berger2001} and Atienza \cite{Atienza2006}. Such classifications implied a complex taxonomy of DMMs, as well as a huge search space. The execution time of both the simulator and the GEA became prohibitive and a parallelization was needed \cite{RiscoMartin2010b}. Furthermore, the profiling of the application was obtained overloading both \texttt{malloc()} and \texttt{free()} functions. Such mechanism required the modification and re-compilation of every target application. Thus, in this new framework, with the use of instrumentation code, the re-compilation is no longer needed, which is a great advantage. Thus, we have renewed both the GEA and the simulator improving the previous search space (too complex for GEA) and the simulation performance, reaching a straightforward design that offers better results in affordable computational times (from minutes to several hours, depending on the target application, vs. days or weeks in other approaches \cite{RiscoMartin2009b}). As a consequence, we present in this work a flexible, stable and highly-configurable DMM design framework. By profiling the target application, the DMM simulator can receive the dynamic behavior of the application off-line and evaluate all the aforementioned metrics. As a result, the simulator is integrated into a search mechanism based on grammatical evolution in order to obtain optimum DMMs.


\section{Dynamic memory management}\label{sec:DynamicMemoryManagement}
In this Section, we summarize the main characteristics of dynamic memory management, as well as the new classification of memory allocators, which is later considered in the implementation of the simulator and the grammar definition.

\subsection{Dynamic memory management}
Dynamic memory management basically consists of two separate tasks, i.e., allocation and deallocation. Allocation is the mechanism that searches for a memory block big enough to satisfy the memory requirements of an object request in a given application. Deallocation is the mechanism that returns a freed memory block to the available memory of the system in order to be reused subsequently. In current applications, the blocks are requested and returned in any order. The amount of memory used by the memory allocator grows up when the memory storage space is used inefficiently, reducing the storage capacity. This phenomenon is called fragmentation. Internal fragmentation happens when requested objects are allocated in blocks whose size is bigger than the size of the object. External fragmentation occurs when no blocks are found for a given object request despite enough free memory is available. Hence, on top of memory de/allocation, the memory allocator has to take care of memory usage issues. To avoid these problems, some allocators include splitting (breaking large blocks into smaller ones to allocate a larger number of small objects) and coalescing (combining small blocks into bigger ones to allocate objects for which there are no available blocks of their size). However, the algorithms that implement these features usually reduce performance due to the additional data structures needed to keep track of the free and used blocks \cite{Jones2012}.

There exist several general-purpose DMMs. Here we briefly describe two of them, the Kingsley DMM \cite{Wilson1995} and the Lea DMM \cite{Lea}. Our optimized DMMs are compared with them because Kingsley and Lea are widely used in both general-purpose and embedded systems, and they are on opposite ends between maximizing performance and minimizing memory usage.

The Kingsley DMM was originally used in BSD 4.2, and current Windows-based systems (both mobile and desktop) apply the main ideas from Kingsley. The Kingsley DMM organizes the available memory in power-of-two block sizes: all allocation requests are rounded up to the next power of two. This rounding can lead to severe memory usage issues, because in the worst case, it allocates twice as much memory as requested. It performs no splitting (breaking large blocks into smaller ones) or coalescing (combining adjacent free blocks). This algorithm is well known to be among the fastest DMMs although it is among the worst in terms of memory usage \cite{Berger2001, Berger2002}.

On the contrary, the Lea DMM is an approximate best-fit DMM that provides mainly low memory usage. Linux-based systems use as their basis the Lea DMM. It presents a different behavior based on the size of the requested memory. For example, small objects (less than 64 bytes) are allocated in free blocks using an exact-fit policy (one linked list of blocks for each multiple of 8 bytes). For medium-sized objects (less than 128Kb), the Lea allocator performs immediate coalescing and splitting in the previous lists and approximates best-fit. For large objects ($\geq$ 128Kb), it uses virtual memory (through \texttt{mmap}).

\subsection{Classification of memory allocators}\label{sec:DmmClassification}
Memory allocators are typically categorized by the mechanisms that manage the lists of free blocks (free lists). These mechanisms include segregated free lists, simple segregated storage, segregated fit, exact segregated fit, strict segregated fit, and buddy systems.

\begin{figure}[!t]
\centering
\includegraphics[width=0.55\columnwidth]{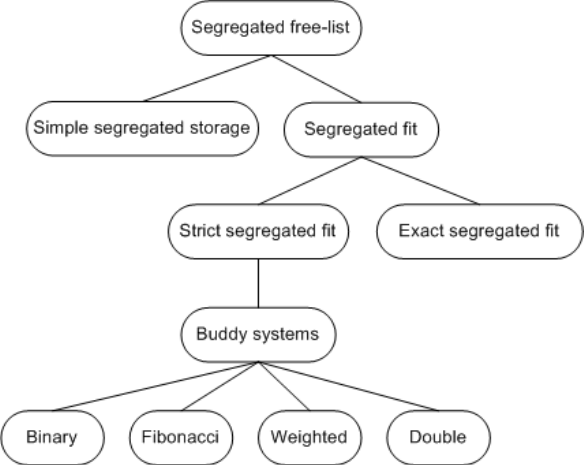}
\caption{Classification of memory allocators.}
\label{fig:DmmClassification}
\end{figure}

\figurename~\ref{fig:DmmClassification} shows the classification of memory allocators provided in \cite{Wilson1995}. To easily describe the search space in a grammar definition, we have reorganized these memory allocator mechanisms in a hierarchy. In the following we briefly describe these allocator mechanisms.

A segregated free-list allocator divides the free list into several subsets, according to the size of the free blocks. A freed or coalesced block is placed on the appropriate list. An allocation request is served from the appropriate list. This class of mechanism usually implements a good fit or best fit policy.

Simple segregated storage is a segregated free-list allocation mechanism which divides the storage into pages or other areas, and only allocates objects of a single size, or small range of sizes, within each area. This approach makes allocation fast and avoids headers, but may lead to high external fragmentation, as unused parts of areas cannot be reused for other object sizes.

Segregated fit is another variation of the segregated free-list class of allocation mechanisms. It maintains an array of free lists, each list holding free blocks of a particular range of sizes. The manager identifies the appropriate free list and allocates from it (often using a first-fit policy). If this mechanism fails, a larger block is taken from another list splitting it accordingly.

Strict segregated fit is a segregated fit allocation mechanism which has only one block size on each free list. A requested block size is rounded up to the next provided size, and the first block on that list is returned. The sizes must be chosen so that any block of a larger size can be split into a number of smaller sized blocks.

Exact segregated fit is a segregated fit allocator, which has a separate free list for each possible block size. The array of free lists may be represented sparsely. Large blocks may be treated separately. The details of the mechanism depend on the distribution of sizes between free lists.

Buddy systems are special cases of strict segregated fit allocators, which make splitting and coalescing fast by pairing each block with a unique adjacent buddy block. To this end, an array of free lists exists, namely, one for each allowable block size. Allocation rounds up the requested size to an allowable size and allocates from the corresponding free list. If the free list is empty, a larger block is selected and split. A block may only be split into a pair of buddies. A block may only be coalesced with its buddy, and this is only possible if the buddy has not been split into smaller blocks. Different sorts of buddy system are distinguished by the available block sizes and the method of splitting. They include binary buddies (the most common type), Fibonacci buddies, weighted buddies, and double buddies \cite{Wilson1995}.


\section{DMM simulator design}\label{sec:TheSimulator}
The global DMM optimization process needs an external simulator to evaluate every candidate DMM generated by the GEA (see \figurename~\ref{fig:Framework3}). In this section we motivate and describe our DMM simulator as well as we outline its design goals.

As introduced in Section \ref{sec:Introduction}, several existing libraries allow the implementation of both general-purpose and custom DMMs. However, exploration techniques cannot be easily applied. Indeed, each custom design must be implemented, compiled and validated against a target application; hence, even if the DMM library is highly modular, this is a very time-consuming process. Thus, a simulator can greatly help in such optimization by being part of a higher optimization module that allows system designers to evaluate (in terms of performance and memory usage) a candidate DMM for the target application. Thus, the desired design goals for the development of this DMM exploration framework are:

\begin{itemize}
\item Efficiency: since the simulator needs to collaborate with the search algorithm, the DMM simulator must improve the execution time of a real DMM.
\item Flexibility: software engineers must be able to simulate any DMM as a composition of single memory allocators. Thus, the parameters of each allocator should be highly configurable.
\end{itemize}

The global design of our software platform is constituted by a DMM simulation library, an optimization library (based on Grammatical Evolution) and a basic \textit{Graphical User Interface (GUI)}. We also have implemented both the Kingsley and Lea memory allocators \cite{Wilson1995}. All the source code is publicly available in \cite{PABA}.

In order to work with our simulator, the engineer must start with a profiling phase of the target application. To this end, we have used a tool called Pin \cite{Luk2005}. Pin's instrumentation is easy to work with. It allows the designer to insert calls to instrumentation at arbitrary locations in the target application source code. The Pin distribution includes many sample architecture-independent Pin-tools including profilers, cache simulators, trace analyzers, and memory bug checkers. \ref{app:PinConf} details the source code and configuration needed to obtain a profiling report of the target application. Note that the software engineer does not need the source code of the target application to obtain the profiling report.

Once obtained the profile, the next step consists of: (a) performing an automatic exploration as described in \figurename~\ref{fig:Framework3}, or (b) selecting the best design among a predefined set of DMM candidates. As a consequence, the composition of a DMM candidate follows a straightforward process for the designer. For instance, one common way of improving memory allocation performance is allocating all objects from a highly-used class using a per-class pool of memory. Because all these objects are of the same size, memory can be managed by a simple singly-linked free-list. Thus, programmers often implement these per-class allocators in C++ by overloading the \texttt{new} and \texttt{delete} operators for the class. In this regard, \ref{app:SimConf} provides the details to configure custom DMMs in two ways: (1) programming, composing DMMs with a few lines of code, or (2) via a \textit{Graphical User Interface (GUI)}, which allows us to select a DMM in a range of preconfigured ones. \ref{app:SimConf} also shows how we can analyze the profiling report to check the different block sizes de/allocated in the target application. We can also study the different classes used in the application as well as their sizes. 

On each simulation run, several relevant metrics are computed, such as the number of de/allocations, splittings, coalescings, performance, memory usage, memory accesses, etc. See \ref{app:SimConf} for more details.

Following the scheme provided in \figurename~\ref{fig:Framework3} we now proceed to describe the other module involved in our design framework: the optimization of DMMs using a Grammatical Evolution Algorithm.


\section{Grammatical evolution applied to DMM optimization}\label{sec:TheGrammar}

\emph{Grammatical Evolution (GE)} (e.g., \cite{ONeill2003}, \cite{Brabazon2006}, \cite{Dempsey2007}, \cite{ONeill2001}, \cite{Brabazon2008}) is a grammar-based form of \emph{Genetic Programming (GP)} \cite{Poli2008}. It combines principles from molecular biology to the representational power of formal grammars. GE has a rich modularity, which provides a unique flexibility. This makes possible to use alternative search strategies (evolutionary, deterministic or some other approach), changing the behavior of the algorithm by merely changing the grammar supplied. Since we use a grammar to define the structures that are generated by GE, we can easily modify the output structures. To this end, we must simply edit the plain text grammar. This is one of the advantages that makes the GE approach so attractive. The characteristics of the GE genotype-phenotype mapping also means that we can operate in traditional integer or binary chromosomes (instead of operating on solution trees, as in standard GP). When solving a problem with GE, a suitable \emph{Backus Naur Form (BNF)} grammar definition must initially be defined. The BNF can be either the specification of an entire language or, perhaps more usefully, a subset of a language geared towards the problem at hand. In a simulation run, GE can theoretically evolve programs in any language described by a BNF.

A grammar can be represented by the tuple $\left\{N, T, P, S\right\}$, where $N$ is the set of non-terminals, $T$ the set of terminals, $P$ a set of production rules that maps the elements of $N$ to $T$, and $S$ is a start symbol that is a member of $N$. When there are a number of productions that can be applied to one element of $N$, the choice is delimited with the ``$|$'' symbol.

\begin{figure}[!t]
\begin{lstlisting}[basicstyle=\scriptsize\ttfamily,breaklines=true,frame=tb]
N = {<DynamicMemoryManager>, <Allocators>, 
    <AllocatorSize>, <AllocatorMaxSize>, 
    <AllocatorClass>, <Size>, <MaxSize>, 
    <AllowSplitting>, <AllowCoalescing>, 
    <DataStructure>, <AllocationMechanism>, 
    <AllocationPolicy>}
T = {SegregatedFreeList, SimpleSegregatedStorage, ExactSegregatedFit, BuddySystemBinary, BuddySystemFibonacci, 4, 7, 8, (block sizes ommited here for the sake of space)... , 7832240, true, false, SLL, DLL, BTREE, FIRST, BEST, EXACT, FIFO, LIFO}
S = <DynamicMemoryManager>
I    <DynamicMemoryManager> ::= <Allocators>
II   <Allocators>           ::= <AllocatorSize>
                          |<AllocatorMaxSize>
III  <AllocatorSize>        ::= <AllocatorClass>
                           <AllowSplitting>
                           <AllowCoalescing>
                           <Size>
                           <DataStructure>
                           <AllocationMechanism>
                           <AllocationPolicy>
                           <Allocators>
IV   <AllocatorMaxSize>     ::= <AllocatorClass>
                           <AllowSplitting>
                           <AllowCoalescing>
                           <MaxSize>
                           <DataStructure>
                           <AllocationMechanism>
                           <AllocationPolicy>
V    <AllocatorClass>    ::= SegregatedFreeList
                       |SimpleSegregatedStorage
                       |ExactSegregatedFit
                       |BuddySystemBinary
                       |BuddySystemFibonacci
VI   <Size>              ::= 4|7|8|12|14|15|16|17|18|19|20|21|23|24|25|26|28|31|32|36|40|44
                       |48|52|56|60|64|72|76|79|80|84|88|92|96|100|104|108|112|116|120|128
                       |132|136|140|144|152|160|168|172|176|180|184|188|192|200|202|208|212
                       |216|220|224|228|232|236|240|248|252|256|264|276|280|284|288|300|303
                       312|316|320|324|336|348|352|356|360|380|384|396|400|424|432|440|444
                       |448|457|480|492|496|500|504|512|525|528|536|540|556|564|576|584|600
                       |608|617|624|626|630|631|632|648|660|672|692|713|720|745|756|768|772
                       |784|800|808|820|...|820840|995732|1011432|1034632|1274436|1295112
                       |1296000|1616064|1916188|1947512|1991464|2548872|2760556|2804912
                       |3832376|3916120|3979304|5521112
VII  <MaxSize>              ::= 7832240
VIII <AllowSplitting>       ::= true|false
IX   <AllowCoalescing>      ::= true|false
X    <DataStructure>        ::= SLL|DLL|BTREE
XI   <AllocationMechanism>  ::= FIRST|BEST|EXACT
XII  <AllocationPolicy>     ::= FIFO|LIFO
\end{lstlisting}
\caption{Grammar file generated with the DealII profiling report. More than 200 block sizes have been omitted for simplicity. The only differences between grammar files generated for different benchmarks are both the \texttt{Size} and \texttt{MaxSize} non-terminals.}
\label{fig:LinsayGrammar}
\end{figure}

We have defined a grammar to optimize DMMs using GE. Such grammar follows the DMM's classification described in Section~\ref{sec:DmmClassification}. Since the grammar depends on the different block sizes managed by the target application, we have developed a tool that generates a grammar file from a given profiling report. \figurename~\ref{fig:LinsayGrammar} shows the grammar generated for DealII, one of the benchmarks analyzed in this paper, which is described in Section \ref{sec:SetUp}. As \figurename~\ref{fig:LinsayGrammar} shows, the grammar contains 12 non-terminals, 31 terminals, and 12 production rules.

Following the production rules, a DMM is a list of several allocators (see production rules I and II in \figurename~\ref{fig:LinsayGrammar}). There are two kinds of allocators (rule II): \texttt{AllocatorSize}, which is followed by more allocators, and \texttt{AllocatorMaxSize}, which is the last allocator in the chain. In this sense, production rules III and IV are equivalent with the only difference that an \texttt{AllocatorMaxSize} is employed to manage block sizes belonging to the biggest range.

As rules III and IV suggest, the general behavior of the allocator is defined in its allocator class, which specifies the way in that list (or arrays) of blocks are distributed. We have developed up to five different classes (see rule V). Both \texttt{AllocatorSize} or (\texttt{AllocatorMaxSize}) production rules determine if the corresponding allocator allows splitting and/or coalescing. Every allocator is responsible to manage block sizes in the range from \texttt{Size} bytes to the \texttt{Size} defined in the following allocator. If the allocator is the last one, i.e. and \texttt{AllocatorMaxSize}, then it manages block sizes in the range $(\textrm{Size}, \textrm{MaxSize}]$. Both \texttt{Size} and \texttt{MaxSize} production rules (VI and VII in \figurename~\ref{fig:LinsayGrammar}) depend on the profiling report of the target application, because each application manages its own range of block sizes.

In this regard, we have observed that different profiling reports coming from the same application are quite similar in both block sizes and number of mallocs and frees. This occurs even if the application is purely interactive (like video games, that need human input), as happens with many multimedia applications. As a consequence, only one profiling report is needed to obtain an optimized DMM per application.

In addition to block sizes, rules III and IV also declare, for a given allocator, which data structure is used to store the blocks (three data structures have been declared, see rule X), the allocation mechanism (first, best and exact fits are allowed, see rule XI) and the allocation policy (FIFO and LIFO, see rule XII). 

A GE's individual uses a variable-length encoding scheme where each gene holds an integer value that will be mapped to previously labeled production rules of a given BNF by the decoding process. The genotype is used to map the start symbol, as defined in the Grammar, onto terminals by reading codons to generate a corresponding integer value. 

Consider the genome example of Figure \ref{fig:GenotypeExample}, which has 18 genes with integer values, and the grammar in \figurename~\ref{fig:LinsayGrammar}. In order to decode the genome, the modulus operation is applied between the gene value and the number of options of each production rule. In this regard, the maximum integer value admitted for each gene is the maximum number of options in a rule of the given grammar.

\begin{figure*}[ht]
\begin{center}
	\begin{tabular}{|c|c|c|c|c|c|c|c|c|c|c|c|c|c|c|c|c|c|}
	\hline
		401 & 213 & 8 & 151 & 77 & 2 & 34 & 60 & 300 & 114 & 205 & 7 & 2 & 122 & 183 & 197 & 49 & 136 \\
\hline
	\end{tabular}
\end{center}
\caption{A GE individual's genome.}
\label{fig:GenotypeExample}
\end{figure*}

Therefore, the decoding process reads the first gene, namely, 401. Since there is only one production rule headed by \texttt{<DynamicMemoryManager>} (group \textbf{I}), the selected one is the production 0 (401 mod 1 = 0) (\texttt{<DynamicMemoryManager> ::= <Allocators>}). Next, the second gene is read and its value (213), after the modulus operation (213 mod 2), results in 1; therefore, the production \texttt{<Allocators> ::= <AllocatorSize>} is picked up. Next, there are eight productions pointed to by the symbol \texttt{<AllocatorSize>}; thus, the genes 8, 151, 77, 2, 34, 60, 300 and 114 are picked up. After the modulus operation, it results in:
\begin{itemize} 
\item AllocatorClass (8 mod 5 = 3): BuddySystemBinary
\item AllowSplitting (151 mod 2 = 1): false
\item AllowCoalescing (77 mod 2 = 1): false
\item Size (2 mod 419 = 2): 8
\item DataStructure (34 mod 3 = 1): DLL
\item AllocationMechanism (60 mod 3 = 0): FIRST
\item AllocationPolicy (300 mod 2 = 0): FIFO
\item Allocators (114 mod 2 = 0): AllocatorMaxSize
\end{itemize}

So far the decoded expression is formed by one allocator, using a BuddySystemBinary behavior. It does not allow us splitting and/or coalescing. In addition, the decoded DMM until now manages block sizes in the range of $(0,8]$ bytes. The data structure used to store blocks is a doubly-linked list. Finally, the allocation mechanism and allocation policy are first fit and first in first out, respectively.

Following the same procedure for the rest of the individual, we decode the next seven genes (205, 7, 2, 122, 183, 197 and 49), which after the modulus operation, result in:

\begin{itemize} 
\item AllocatorClass (205 mod 5 = 0): SegregatedFreeList
\item AllowSplitting (7 mod 2 = 1): false
\item AllowCoalescing (2 mod 2 = 0): true
\item MaxSize (122 mod 1 = 0): 7832240
\item DataStructure (183 mod 3 = 0): SLL
\item AllocationMechanism (197 mod 3 = 0): FIRST
\item AllocationPolicy (49 mod 2 = 1): LIFO
\end{itemize}

\begin{table}[ht]
\caption{Resultant Dynamic Memory Manager encoded in \figurename~\ref{fig:GenotypeExample}.}
\begin{center}
\begin{tabular}{ccc}
\hline
\multicolumn{3}{c}{BuddySystemBinary, split=false, coalesce=false} \\
Data Structure & Mechanism(Policy) & Range (bytes) \\
DLL & FIRST(FIFO) & (0, 1] \\
DLL & FIRST(FIFO) & (1, 2] \\
DLL & FIRST(FIFO) & (2, 4] \\
DLL & FIRST(FIFO) & (4, 8] \\
\hline
\multicolumn{3}{c}{SegregatedFreeList, split=false, coalesce=true} \\
Data Structure & Mechanism(Policy) & Range (bytes) \\
SLL & FIRST(LIFO) & (8, 7832240] \\
\hline
\end{tabular}
\end{center}
\label{tab:GenotypeExample}
\end{table}

The decoded DMM after the 17th gene is represented in \tablename~\ref{tab:GenotypeExample}. Such DMM is composed by two allocators, both defined in terms of the decoded individual depicted in \figurename~\ref{fig:GenotypeExample}. Then, as shown in \figurename~\ref{fig:Framework3}, the simulator is continuously receiving best individuals in the optimization process. Hence, it decodes the individual and simulates the resulting DMM using the profiling report, giving its fitness in terms of performance and memory usage.

An important advantage of the grammatical evolution representation shown is that it uses a linear genome. Therefore, GE can directly use all standard genetic algorithm operators. Furthermore, because of the simplicity of the linear representation, computing implementations of GE are relatively simple to deploy.

Indeed, in the previous example the decoding process terminated without translating all genes. This occurs because in GE the genetic operations do not know about the semantics of a genome until it is decoded, so the decoding process frequently ends up with a complete expression (final) without traversing the entire genome. Hence, this exploration process has a short exploration time.

On the other hand, we can find a potentially more serious concern with GE, which is the situation of having redundant genes. In the decoding process, it could happen that a genome does not have sufficient genes to create a complete phenotype, i.e., there still exist non-terminals in the chromosome. In this case the \emph{wrap} operator is applied. It consists of returning the reading process head back to the first codon in the chromosome. Even after the first wrapping, the mapping process could be incomplete and would carry on indefinitely unless terminated. This occurs because non-terminals are being mapped recursively by several production rules. This individual should be marked as invalid. For this reason an upper limit on the number of wrappings is set in our optimization approach for DMMs. Therefore, during the mapping process, starting from the left-hand side of the genome, we generate integer values and use them to select rules from the BNF grammar, until one of the following situations arise:

\begin{enumerate}
	\item A complete DMM is generated. It occurs when all the non-terminals are transformed into elements from the terminal set of the BNF grammar.
	\item The end of the genome is reached and then the wrapping operator is called. This results in the continuation of the genome reading starting from the left hand side of the genome once again. The reading of the codons will then continue, unless an upper limit representing the maximum number of wrappings has occurred during this indidual's mapping process.
	\item In the case that the limit of the number of wrappings has occurred and the individual is still incompletely mapped, the mapping process is stopped, and the individual is assigned the highest fitness value (assuming minimization).
\end{enumerate}

Our grammar is complete enough to implement any well-known DMMs and to explore custom DMM implementations for real-life multimedia applications (see Section~\ref{sec:Experiments} for more details). Regarding hardware platforms, it can be applied to any computer platform where the use of C++ is allowed.


\section{Experimental methodology}\label{sec:SetUp}

In this section we describe the set of benchmarks used, the fitness function employed and the general configuration of our optimization algorithm.

\subsection{Benchmarks}
As stated above, we study the performance and memory usage implications of building general-purpose and custom allocators using both the simulator and our GEA proposal. To evaluate both metrics, we have selected a number of memory-intensive programs \cite{SPECpu2006}, some of them executed with different inputs, as listed in \tablename~\ref{tab:Benchmarks} and next described.

\begin{table*}[ht]
	\caption{Statistics for the analyzed memory-intensive benchmarks}
	\centering
	\begin{tabular}{lrrrrr}
\hline
\multicolumn{6}{c}{\textbf{Memory-Intensive Benchmark Statistics}} \\
\hline
Benchmark     & Objects & Total memory (bytes) & Max in use (bytes) & Average size (bytes) & Memory ops \\
\hline
hmmer         				& 5000101	& 2780155771		& 66074  		& 556.01		& 10020365 \\
dealII								& 4926536	& 451183299			& 33747256	& 91.58	  	& 10251022 \\
soplex								& 6188584	& 4537088927228	& 13273944	& 733138.45	& 12377148 \\
calculix							& 4054631	& 19675406468		& 92335483 	& 4852.57   & 10432894 \\
gcc/200								& 4810046	& 4568294683		& 163997226	& 949.74    & 10875348 \\
gcc/expr2							& 3363661	& 6246880022		& 531906484 & 1857.16   & 7632761 \\
gcc/c-typeck					& 2786057	& 6040320719		& 408900497 & 2168.05   & 6781339 \\
perl/split						& 5534308	& 1526190192		& 140256527 & 275.76    & 10900336 \\
perl/diff							& 5321169	& 54544092			& 17209415  & 10.25     & 10544463 \\
perl/spam							& 5278732	& 213309132			& 27784844  & 40.40     & 10320179 \\
\hline
	\end{tabular}
	\label{tab:Benchmarks}
\end{table*}

\textit{hmmer} performs sensitive searches in a gene sequence database, by applying the statistic models developed in profile hidden Markov models. These tecniques are widely used in computational biology to search for patterns in DNA sequences and in the protein sequence analysis. 

\textit{dealII} is an application that uses the C++ library named deal.II. This library allows the development of modern adaptive finite elements algorithms in several space dimensions using, among others aspects, sofisticated error estimations and adaptive meshes. The library provides a modern interface to the complex data structures and algorithms required. Also, programs can be developed to be independent of the space dimension, without great penalties on run-time and memory consumption.

\textit{soplex} is an application which solves a linear program applying the Simplex algorithm. A linear program is given by a sparse matrix $A$ of $m$ by $n$, and a right hand side vector $b$ of dimension $m$ and an objective function coefficent vector $c$ of dimension $n$. The goal is usually to find the vector $x$ that minimizes $c$ with $Ax \leq b$ and $x \geq 0$.

\textit{calculix} is applied in structural mechanics and it is based on Calculix, a software that uses clasical theory of finite elements. It is used for linear and non linear three dimensional structural applications and can be used to solve problems related with the design of bridges, building, earthquake resistance, resonance phenomena, etc.

\textit{gcc} is a C language optimizing compiler based on gcc version 3.2. It adds optimized code generation for an AMD processor. The inlining of heuristics have been slightly modified in order to add code more typical of current compiler usage. Also, gcc would spend more time analyzing the source code inputs and use more memory. In addition, we have profiled this application with three different inputs: 200.i comes from a previous version of the floating point SPEC2000 benchmark (SPECfp2000) 200.sixtrack (gcc/200). Expr2.i comes from the source of 403.gcc (gcc/expr2), as c-typeck.i does (gcc/c-typeck).

\textit{perlbench} is a shortened version of PERL, a scripting programming language where OS-specific features have been removed. The workload consists of three scripts with several third-party modules. The first workload is the open source spam checking software SpamAssasin, used to score a couple of known corpora of both spam and ham. MhonArc is a popular freeware email-to-HTML converter, where email messages are generated randomly and converted to HTML. Lastly, a slightly-modified version of specdiff script, as part of the CPU2006 tool suite.  Thus, we have profiled this application with three different inputs: splitmail, based on the splitmail program that takes an email message and break it up into smaller pieces (perl/split); diffmail, which is a differentiated message delivery architecture to control spam (perl/diff); and checkspam, which is a PERL script to check both spam and ham using the SpamAssasin module (perl/spam). 

To perform the profiling of the applications, we applied the Pin tool, as described in Section \ref{sec:TheSimulator}, and we run all the applications on an Intel Core 2 Quad processor Q8300 system with 4 GB of RAM, under Windows 7. This task was performed just once within the proposed DMM exploration methodology.

\tablename~\ref{tab:Benchmarks} includes the number of objects allocated and their average size. The applications' memory usages range from just 64.5 KB (for hmmer) to over 507.27 MB (for gcc/expr2). For all the programs, the ratio between total allocated memory and the maximum amount of memory in use is large. In addition, the number of memory operations is also large. Thus, all the proposed benchmarks highly rely on the dynamic memory subsystem. In the next step we simulate five general-purpose allocators, as well as perform an automatic exploration of feasible DMMs using grammatical evolution over all the tested benchmarks.

\subsection{Fitness function}
In order to select the best possible DMM among a set of candidates, the optimization algorithm must use a well-defined fitness function. In this work, we have used a weighted sum of execution time and memory usage, normalized to both Kingsley and Lea DMMs, the fastest and the more memory efficient, respectively. Thus, the fitness function is given by:
\begin{equation}\label{eq:Fitness}
F = 0.5 \cdot \frac{T}{T_\mathrm{KNG}} + 0.5 \cdot \frac{M}{M_\mathrm{LEA}}
\end{equation}
\noindent where $T$ and $M$ are the execution time and memory used by the DMM being evaluated, and $T_\mathrm{KNG}$, $M_\mathrm{LEA}$ are the execution time and memory usage of Kingsley and Lea DMMs, respectively. It is worth noting that we define memory usage as the high water-mark of memory requested from the virtual memory.

\subsection{Algorithm configuration}
To implement our GE algorithm, we have used JECO \cite{PABA}, a well-known optimization library developed in Java. With the parameters given below, we have set 250 generations and 100 individuals because is the minimal number of evaluations needed by GEA to reach the best DMM in the Soplex benchmark, which is the biggest one ($12\times 10^6$ memory operations, as \tablename~\ref{tab:Benchmarks} shows). Obviously, with the number of evaluation fixed, the smaller the size of the profiling report (with respect to Soplex), the higher the quality of solutions. Thus, we have selected the following parameters:
\begin{itemize}
\item Population size: 100
\item Elite size: 10
\item Replacement type: generational
\item Number of generations: 250
\item Selection mechanism: Tournament
\item Tournament size: 2
\item Probability of crossover: 0.80
\item Fixed point crossover: yes
\item Probability of mutation: 0.02
\item Maximum number of wraps: 3
\end{itemize}

Nevertheless, it should be noted that in our methodology the simulator can obtain the fitness of a generic DMM in a couple of seconds (the Kingsley DMM, for example) and continue with the optimization process until a given fitness have been reached (75\% of Kingsley's, for example).


\section{Results}\label{sec:Experiments}

We simulated the benchmarks described in the previous section by considering five general-purpose allocators: the Kingsley allocator (labeled as KNG in the following figures), the Doug's Lea allocator (labeled as LEA), a buddy system based on the Fibonacci allocation algorithm (labeled as FIB), a list of 10 segregated free-lists (S10), and an exact segregated free list allocator (EXA). Finally, we compared the results with the custom DMM obtained with our proposed automatic exploration process (labeled as GEA). To find a solution to our multiobjective optimization problem, we construct the single aggregate objective function shown in Equation (\ref{eq:Fitness}). In addition, to prove that the algorithm is not trapped into a local optima, we have extended the number of generations for both hmmer and soplex benchmarks to $5000$ (labeled as GEA*). All the results obtained are averaged over 10 trials.

To create the custom DMMs, we have followed the proposed methodology flow (\figurename~\ref{fig:Framework}). We first profiled the behavior of the application under study using the Pin tool. Our Grammar Generator tool completes the grammar file (as described in Section \ref{sec:TheGrammar}), filling in the different block sizes demanded by the application, which are initially unknown. Finally, we execute our grammatical evolutionary algorithm. 

\begin{figure*}[t!]
\centering
\subfigure[Global fitness]{
\includegraphics[width=0.95\textwidth]{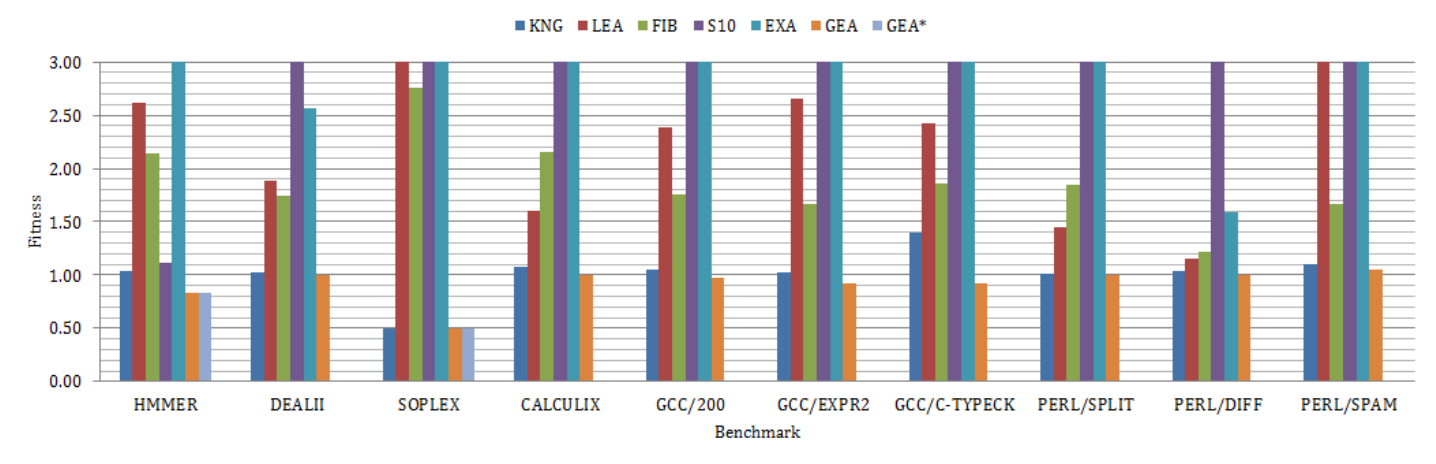}
\label{fig:ResultsFitness}
}
\subfigure[Execution time]{
\includegraphics[width=0.95\textwidth]{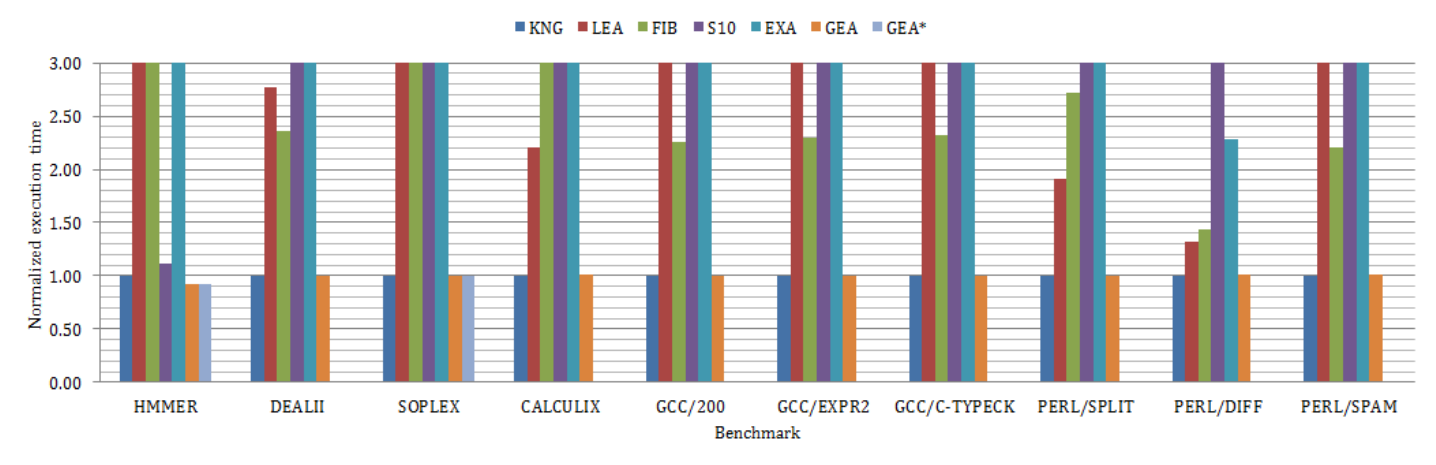}
\label{fig:ResultsPerformance}
}
\subfigure[Memory usage]{
\includegraphics[width=0.95\textwidth]{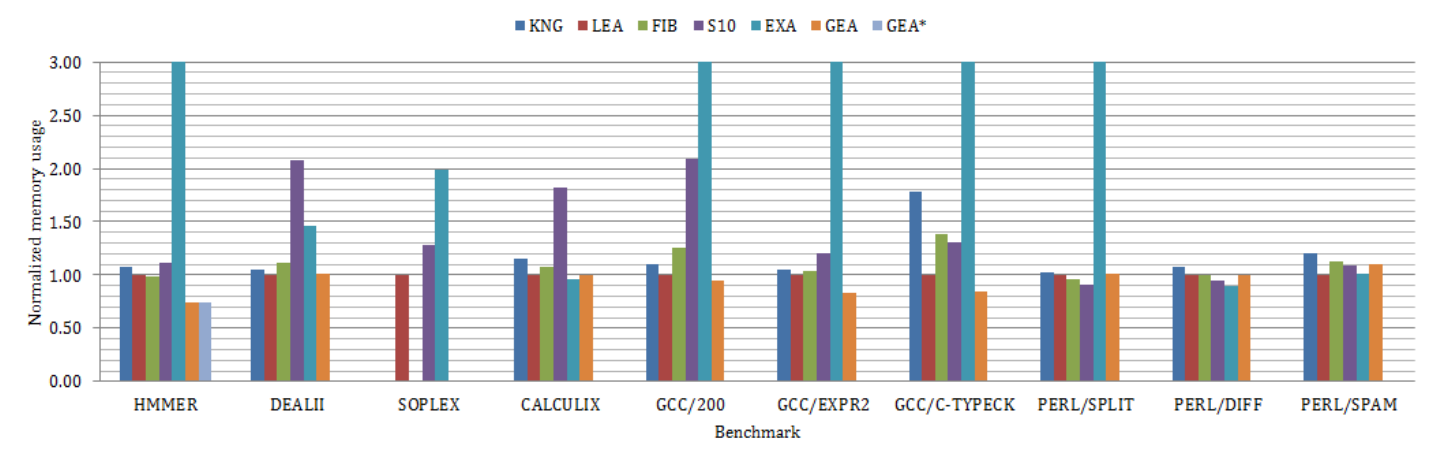}
\label{fig:ResultsMemory}
}
\caption{Global fitness $F$, execution time $\frac{T}{T_\mathrm{KNG}}$ and memory usage $\frac{M}{M_\mathrm{LEA}}$ for the six benchmarks managed by six DMMs.}
\label{fig:Results}
\end{figure*}

In \figurename~\ref{fig:Results} we present our experimental results. To better scale the values, we have limited the upper bound of the bar plots to $3$ ($F_\mathrm{EXA}=2763.79$ in soplex, for instance). \figurename~\ref{fig:ResultsFitness} depicts the global fitness reached by all the proposed DMMs, given by Eq. (\ref{eq:Fitness}). Figs. \ref{fig:ResultsPerformance} and \ref{fig:ResultsMemory} show each one of the components of the global fitness, i.e., normalized execution time $\frac{T}{T_\mathrm{KNG}}$ and normalized memory usage $\frac{M}{M_\mathrm{LEA}}$, respectively. In all the experiments performed, the standard deviation was lower than $0.04$.

\tablename~\ref{tab:PercentageAvg} shows the improvement (in average) obtained by GEA in global fitness, performance and memory use, respectively, with respect to the six benchmarks. As can be seen, GEA outperformed all the five general purpose DMMs in all cases, reaching improvements up to 90.78\% in the global fitness.

\begin{table}[ht]
\caption{Averaged percentage of improvent. GEA vs. KNG, LEA, FIB, S10 and EXA.}
\centering
\begin{tabular}{ccccccc}
\hline
 & KNG & LEA & FIB & S10 & EXA & Average \\
\hline
Fitness ($=100 \times \frac{F_\mathrm{*}-F_\mathrm{GEA}}{F_\mathrm{*}}$) & 9.13\% & 62.52\% & 51.81\% & 86.88\% & 86\% & 59.27\% \\
Performance ($=100 \times \frac{T_\mathrm{*}-T_\mathrm{GEA}}{T_\mathrm{*}}$) & 1.17\% & 72.44\% & 62.62\% & 85.74\% & 90.78\% & 62.55\% \\
Memory ($=100 \times \frac{M_\mathrm{*}-M_\mathrm{GEA}}{M_\mathrm{*}}$) & 16.03\% & 23.14\% & 15.08\% & 38.88\% & 59.96\% & 30.62\% \\
\hline
\end{tabular}
\label{tab:PercentageAvg}
\end{table}

As a global analysis of general-purpose DMMs, we can see in both \figurename~\ref{fig:ResultsFitness} and \tablename~\ref{tab:PercentageAvg} that Kingsley is an excellent allocator. It is clearly the best general-purpose DMM in terms of performance (see \figurename~\ref{fig:ResultsPerformance}). This great performance is slighty damaged by the memory usage (see \figurename~\ref{fig:ResultsMemory}), which is the feature where KNG is outperformed, mainly by LEA, in nine of the ten cases. As a consequence, LEA is a good candidate in order to reduce memory usage. It is worth noting than LEA was widely outperformed by Kingsley in soplex. The buddy system based on Fibonacci (FIB) always shows an intermediate behavior in five of the six benchmarks. The other two general-purpose DMMs, labeled S10 and EXA, present a different behavior, depending on the benchmark. For example, examining the global fitness (\figurename~\ref{fig:ResultsFitness}), the ten segregated lists (S10) is a good allocator for hmmer, but it is the worst choice for dealII, calculix, gcc and perlbench. Finally, the exact segregated list (EXA) did not respond well, being the worst choice for hmmer and soplex.

We now analyze the results obtained by both GEA and GEA*, examining each metric independently. To this end we also show in \tablename~\ref{tab:Percentage} the percentage of improvement reached by our exploration algorithm, compared to the Kingsley allocator. After a first view to the global fitness (see also \figurename~\ref{fig:ResultsFitness}) we may conclude that our custom DMM is the best choice in all the six benchmarks (ten configurations), giving improvements of 20.08\%, 2.31\%, 0.17\%, 6.59\%, 7.21\%, 10.55\%, 33.35\%, 1.11\%, 3.61\% and 4.30\% in the cases of hmmer, dealII, soplex, calculix, gcc (200, epxr2, c-typeck), and perlbench (split, diff, spam), respectively. It is worth noting that the results was exactly the same (in average) using $5000$ generations instead of $250$ (see hmmer* and soplex* in \tablename~\ref{tab:Percentage}). It makes sense since both GEA and GEA* finished quite close to the Kingsley's performance, which is difficult to improve. Finally, the wall-clock time needed to evaluate 250 generations of 100 individuals varies from a couple of hours in the case of hmmer to 9-10 hours in the case of soplex.

\begin{table}[ht]
\caption{Percentage of improvent. GEA (GEA*) vs. Kingsley}
\centering
\begin{tabular}{cccc}
\hline
 & $100 \times \frac{F_\mathrm{KNG}-F_\mathrm{GEA}}{F_\mathrm{KNG}}$ & $100 \times \frac{T_\mathrm{KNG}-T_\mathrm{GEA}}{T_\mathrm{KNG}}$ & $100 \times \frac{M_\mathrm{KNG}-M_\mathrm{GEA}}{M_\mathrm{KNG}}$ \\
\hline
hmmer & 20.08\% & 8.33\% & 31.04\% \\
hmmer* & 20.08\% & 8.33\% & 31.04\% \\
dealII & 2.31\% & 0.11\% & 4.58\% \\
soplex & 0.17\% & 0.08\% & 3.72\% \\
soplex* & 0.17\% & 0.08\% & 3.72\% \\
calculix & 6.59\% & -0.84\% & 13.17\% \\
gcc/200 & 7.21\% & 0.01\% & 14.05\% \\
gcc/expr2 & 10.55\% & -0.37\% & 20.80\% \\
gcc/c-typeck & 33.35\% & -0.49\% & 52.44\% \\
perl/split & 1.11\% & -0.04\% & 2.05\% \\
perl/diff & 3.61\% & -0.58\% & 7.65\% \\
perl/spam & 4.30\% & -0.54\% & 8.14\% \\
\hline
\end{tabular}
\label{tab:Percentage}
\end{table}

Regarding performance (\figurename~\ref{fig:ResultsPerformance}) the best results in this case are obtained by Kingsley and GEA. Note that in 6 of 10 cases, GEA is up to 0.84\% worse than Kingsley. It is not relevant, because Kingsley is a highly optimized  allocator for performance, while tends to perform worse than other managers regarding to memory footprint \cite{Atienza2006a}. Then, Lea, the Fibonacci-based buddy allocator, Segregated list and Exact fit, on the contrary, have different behaviors depending on the application (particularly S10 and EXA). Thus, these four allocators are worse than Kingsley and GEA in all conditions.

In terms of memory usage (\figurename~\ref{fig:ResultsMemory}), we can observe that Kingsley is not the best allocator, as it suffers from a high level of internal fragmentation which in turn results in a larger memory utilization. Although Lea is highly optimized in memory usage, we can see that it does not perform well in the case of soplex. This occurs because soplex uses almost all the block sizes in the range $\left[ 1 \ldots 3849000 \right]$ bytes. Then Lea uses big block sizes and internal fragmentation occurs. Once more, the DMM obtained by GEA is the best choice: 31.04\%, 4.58\%, 3.72\%, 13.17\%, 14.05\%, 20.80\%, 52.44\%, 2.05\%, 7.65\%, and 8.14\% better than Kingsley in hmmer, dealII, soplex, calculix, gcc (200, epxr2, c-typeck), and perlbench (split, diff, spam), respectively. As a result, GEA is the best choice.

At this point, we can observe that using our proposed DMM simulator in the memory exploration process enables the study of six different allocation mechanisms starting from a unique previous run of each target application. This point, that comes from joining the profiling task and the simulator features, enables large savings in the dynamic memory optimization exploration, providing great benefit to designers.

We have analyzed the numerical results obtained using our methodology. Next, we describe the structure of the best custom DMMs obtained by GEA for all the ten configurations.

\subsection{hmmer}

\tablename~\ref{tab:hmmer} shows the DMM obtained by both GEA and GEA* for the hmmer benchmark. This benchmark includes 2013 different block sizes, varying from 2 to 9696 bytes (9.47 KB). The custom DMM presents 2 different internal allocators. The first one is a segregated free list with only coalescing, where its unique list of free-blocks covers the range $(2 \ldots 9600]$ bytes, and is implemented as a B-tree with first-fit allocation and first-in-first-out policy. The allowed range by the free list is depicted in the last column of \tablename~\ref{tab:hmmer}. The second allocator follows an exact segregated fit with splitting and coalescing. It contains one free list implemented with a binary tree, with first-fit and last-in-first-out policies, varying from 9.37 to 9.47 KB. Since there are no block sizes between 9600 and 9696 bytes, it is logical to have one exact allocator to store the blocks with the maximum size. In this way, the performance is close to the kinsgley allocator (having only 2 data structures) and the memory is better managed.

\begin{table}[ht]
\caption{Custom DMM map for the hmmer benchmark.}
\begin{center}
\begin{tabular}{ccc}
\hline
\multicolumn{3}{c}{SegregatedFreeList, split=false, coalesce=true} \\
Data Structure & Mechanism(Policy) & Range (bytes) \\
BTREE & FIRST(FIFO) & (0, 9600] \\
\hline
\multicolumn{3}{c}{ExactSegregatedFit, split=true, coalesce=true} \\
Data Structure & Mechanism(Policy) & Range (bytes) \\
BTREE & FIRST(LIFO) & (9600, 9696] \\
\hline
\end{tabular}
\end{center}
\label{tab:hmmer}
\end{table}

\subsection{dealII}

In the same format, \tablename~\ref{tab:dealII} shows the DMM obtained for the dealII benchmark. It includes 419 different block sizes, varying from 4 to 7832240 bytes (7.47 MB). The GEA DMM presents 4 different internal allocators, three of them with both splitting and coalescing. The first one is a binary buddy, where all the free-lists are implemented as single-linked lists with exact allocation and LIFO policy, allocating block sizes in the range from 0 to 8192 bytes. The second allocator follows a segregated free list and contain just one free-list, implemented as a binary tree with a first-fit allocation algorithm with LIFO policy convering the range 8 to 64 KB. This allocator is followed by a binary buddy with just one free-list, implemented as a single-linked list with a first-fit allocation algorithm and FIFO policy. The last allocator is again a binary buddy, this time formed by six binary trees with exact allocation algorithm and FIFO policy. In this case, the 96\% of the mallocs are concentrated in the first allocator that allows the custom DMM to obtain a performace similar to Kingsley. In addition, most of the block sizes are power of two. Thus, the exact-fit algorithm is a good choice both in the first and last allocators.

\begin{table}[ht]
\caption{Custom DMM map for the dealII benchmark.}
\begin{center}
\begin{tabular}{ccc}
\hline
\multicolumn{3}{c}{BuddySystemBinary, split=false, coalesce=false} \\
Data Structure & Mechanism(Policy) & Range (bytes) \\
SLL & EXACT(LIFO) & (0, 1] \\
SLL & EXACT(LIFO) & (1, 2] \\
SLL & EXACT(LIFO) & (2, 4] \\
\multicolumn{3}{c}{$\ldots$} \\
SLL & EXACT(LIFO) & (4096,8192] \\
\hline
\multicolumn{3}{c}{SegregatedFreeList, split=true, coalesce=true} \\
Data Structure & Mechanism(Policy) & Range (bytes) \\
BTREE & FIRST(LIFO) & (8192, 65536] \\
\hline
\multicolumn{3}{c}{BuddySystemBinary, split=true, coalesce=true} \\
Data Structure & Mechanism(Policy) & Range (bytes) \\
SLL & FIRST(FIFO) & (65536, 131072] \\
\hline
\multicolumn{3}{c}{BuddySystemBinary, split=true, coalesce=true} \\
Data Structure & Mechanism(Policy) & Range (bytes) \\
BTREE & EXACT(FIFO) & (131072, 262144] \\
BTREE & EXACT(FIFO) & (262144, 524288] \\
BTREE & EXACT(FIFO) & (524288, 1048576] \\
\multicolumn{3}{c}{$\ldots$} \\
BTREE & EXACT(FIFO) & (4194304, 8388608] \\
\hline
\end{tabular}
\end{center}
\label{tab:dealII}
\end{table}

\subsection{soplex}

\tablename~\ref{tab:soplex} shows the DMM obtained by both GEA and GEA* for the soplex benchmark. Soplex uses 47728 different block sizes, varying from 1 to 3849000 bytes (3.67 MB). Our custom DMM is almost equal to the Kingsley allocator, but with splitting and coalescing that allows us to save memory. In this case, memory usage (in terms of block sizes) is equally distributed in the whole range. This diversification does not allow the optimization algorithm to find patterns in which a particular allocator could improve the internal fragmentation.

\begin{table}[ht]
\caption{Custom DMM map for the soplex benchmark.}
\begin{center}
\begin{tabular}{ccc}
\hline
\multicolumn{3}{c}{BuddySystemBinary, split=true, coalesce=true} \\
Data Structure & Mechanism(Policy) & Range (bytes) \\
DLL & FIRST(FIFO) & (0,4194304] \\
\hline
\end{tabular}
\end{center}
\label{tab:soplex}
\end{table}

\subsection{calculix}

\tablename~\ref{tab:calculix} depicts the DMM obtained for calculix. Calculix uses 939 different block sizes, varying from 4 to 8214296 bytes (7.83 MB). Our custom DMM is formed by 3 internal allocators. The first one is a binary buddy, with splitting and coalescing. It includes 17 singly-linked lists, all of them with first-fit mechanism and FIFO policy. The second and third allocators follows a segregated free list mechanism and both are formed by just one free-list, implemented the first as a binary tree, with first-fit mechanism and FIFO policy, and the second as a doubly-linked list with exact-fit mechanism and FIFO policy. The 99.46\% of the block requests are concentrated in the first allocator. Since the first one is a binary buddy, it approximates the performance to the Kingsley allocator. The other two allocators are optimized for the remaining 0.54\%. It allows our custom DMM to save up to 13.17\% of memory usage with respect to Kingsley (see \tablename~\ref{tab:Percentage}).

\begin{table}[ht]
\caption{Custom DMM map for the calculix benchmark.}
\begin{center}
\begin{tabular}{ccc}
\hline
\multicolumn{3}{c}{BuddySystemBinary, split=true, coalesce=true} \\
Data Structure & Mechanism(Policy) & Range (bytes) \\
SLL & FIRST(LIFO) & (0, 1] \\
SLL & FIRST(LIFO) & (1, 2] \\
SLL & FIRST(LIFO) & (2, 4] \\
\multicolumn{3}{c}{$\ldots$} \\
SLL & FIRST(LIFO) & (32768, 65536] \\
\hline
\multicolumn{3}{c}{SegregatedFreeList, split=false, coalesce=false} \\
Data Structure & Mechanism(Policy) & Range (bytes) \\
BTREE & FIRST(FIFO) & (65536, 177240] \\
\hline
\multicolumn{3}{c}{SegregatedFreeList, split=false, coalesce=false} \\
Data Structure & Mechanism(Policy) & Range (bytes) \\
DLL & EXACT(FIFO) & (177240, 8214296] \\
\hline
\end{tabular}
\end{center}
\label{tab:calculix}
\end{table}

\subsection{gcc (200, expr2 and c-typeck)}

\tablename~\ref{tab:gcc} shows the best three DMMs obtained by GEA for the three configurations of the gcc benchmark. The number of block sizes was different in each case, 13964, 5957, and 8177 for gcc/200, gcc/expr2 and gcc/c-typeck, respectively. The two DMMs obtained for gcc/200 and gcc/expr2 were almost the same (two binary buddies, the second one with splitting and coalescing), whereas the last DMM was partially different, being the second allocator a segregated free list. However, we also found a binary buddy as the second allocator among the 5 best DMMs for gcc/c-typeck (in all the 10 runs), saving a 43.13\% of memory use instead of a 52.44\% with respect to the Kingsley DMM (see \tablename~\ref{tab:Percentage}). Thus, we were able to obtain a uniform custom DMM based on two binary buddies for gcc, independently of the selected input.

\begin{table}[ht]
\caption{Custom DMM map for the gcc benchmark (200, expr2 and c-typeck).}
\begin{center}
\begin{tabular}{ccc}
\hline
\hline
\multicolumn{3}{c}{\textbf{gcc/200}} \\
\hline
\hline
\multicolumn{3}{c}{BuddySystemBinary, split=false, coalesce=false} \\
DLL & FIRST(FIFO) & (0,1] \\
DLL & FIRST(FIFO) & (1,2] \\
DLL & FIRST(FIFO) & (2,4] \\
\multicolumn{3}{c}{$\ldots$} \\
DLL & FIRST(FIFO) & (8K,16K] \\
\hline
\multicolumn{3}{c}{BuddySystemBinary, split=true, coalesce=true} \\
BTREE & FIRST(FIFO) & (16K,32K] \\
\multicolumn{3}{c}{$\ldots$} \\
BTREE & FIRST(FIFO) & (8M,16M] \\
\hline
\hline
\multicolumn{3}{c}{\textbf{gcc/expr2}} \\
\hline
\hline
\multicolumn{3}{c}{BuddySystemBinary, split=false, coalesce=false} \\
DLL & FIRST(FIFO) & (0,1] \\
DLL & FIRST(FIFO) & (1,2] \\
DLL & FIRST(FIFO) & (2,4] \\
\multicolumn{3}{c}{$\ldots$} \\
DLL & FIRST(FIFO) & (8K,16K] \\
\hline
\multicolumn{3}{c}{BuddySystemBinary, split=true, coalesce=true} \\
BTREE & FIRST(FIFO) & (16K,32K] \\
\multicolumn{3}{c}{$\ldots$} \\
BTREE & BEST(FIFO) & (128M,256M] \\
\hline
\hline
\multicolumn{3}{c}{\textbf{gcc/c-typeck}} \\
\hline
\hline
\multicolumn{3}{c}{BuddySystemBinary, split=false, coalesce=false} \\
DLL & FIRST(FIFO) & (0,1] \\
DLL & FIRST(FIFO) & (1,2] \\
DLL & FIRST(FIFO) & (2,4] \\
\multicolumn{3}{c}{$\ldots$} \\
DLL & FIRST(FIFO) & (64K,128K] \\
\hline
\multicolumn{3}{c}{SegregatedFreeList, split=false, coalesce=true} \\
BTREE & BEST(FIFO) & (128K,256M] \\
\hline
\end{tabular}
\end{center}
\label{tab:gcc}
\end{table}

\subsection{perlbench (split, diff and spam)}

Finally, the three DMMs obtained by GEA for the perl benchmark are shown in \tablename~\ref{tab:perl}. This benchmark was also configured with 3 different inputs to check the possibility to obtain the same DMM for all the three configurations. In this case, as \tablename~\ref{tab:perl} shows, the three DMMs were radically different. If we examine the three profilings, we may observe that the demand pattern for the three configurations was quite different. We found 3397, 555 and 1863 different block sizes for perl/split, perl/diff and perl/spam, respectively. The most demanded block size was 105 KB, 9 KB, and 1KB in each case, which gives us and idea of the different nature of these three configurations. Regarding the profilings, perl/split directly starts with big block sizes, whereas perl/diff and perl/spam are more concentrated in small block sizes. Thus, our GE algorithm did not find an optimal DMM for all the three configurations.

\begin{table}[ht]
\caption{Custom DMM map for perlbench (split, diff and spam).}
\small
\begin{center}
\begin{tabular}{ccc}
\hline
\hline
\multicolumn{3}{c}{\textbf{perl/split}} \\
\hline
\hline
\multicolumn{3}{c}{BuddySystemBinary, split=false, coalesce=false} \\
DLL & EXACT(FIFO) & (0,1] \\
DLL & EXACT(FIFO) & (1,2] \\
\multicolumn{3}{c}{$\ldots$} \\
DLL & EXACT(FIFO) & (128K,256K] \\
\hline
\multicolumn{3}{c}{SegregatedFreeList, split=true, coalesce=true} \\
BTREE & BEST(FIFO) & (256K,1M] \\
\hline
\hline
\multicolumn{3}{c}{\textbf{perl/diff}} \\
\hline
\hline
\multicolumn{3}{c}{BuddySystemBinary, split=false, coalesce=false} \\
DLL & FIRST(FIFO) & (0,1] \\
DLL & FIRST(FIFO) & (1,2] \\
\multicolumn{3}{c}{$\ldots$} \\
DLL & FIRST(FIFO) & (64,128] \\
\hline
\multicolumn{3}{c}{BuddySystemFibonacci, split=false, coalesce=false} \\
SLL & FIRST(FIFO) & (128,144] \\
SLL & FIRST(FIFO) & (144,233] \\
SLL & FIRST(FIFO) & (233,377] \\
\multicolumn{3}{c}{$\ldots$} \\
BTREE & BEST(FIFO) & (610,987] \\
\hline
\multicolumn{3}{c}{SegregatedFreeList, split=false, coalesce=true} \\
BTREE & BEST(LIFO) & (987,497K] \\
\hline
\hline
\multicolumn{3}{c}{\textbf{perl/spam}} \\
\hline
\hline
\multicolumn{3}{c}{BuddySystemBinary, split=false, coalesce=false} \\
SLL & FIRST(FIFO) & (0,1] \\
SLL & FIRST(FIFO) & (1,2] \\
\multicolumn{3}{c}{$\ldots$} \\
SLL & FIRST(FIFO) & (256,512] \\
\hline
\multicolumn{3}{c}{SegregatedFreeList, split=true, coalesce=true} \\
BTREE & BEST(FIFO) & (512,8K] \\
\hline
\multicolumn{3}{c}{SimpleSegregatedStorage, split=false, coalesce=false} \\
BTREE & BEST(FIFO) & (8K,128K] \\
\hline
\end{tabular}
\end{center}
\label{tab:perl}
\end{table}

To conclude, GEA was able to find optimal DMMs for each of the six benchmarks. In the case of gcc, we were able to find a common DMM for different configurations. However, in the case of the perl benchmark it was not possible, due to the different nature of this benchmark whose behavior strogly depends on the given input. In addition, all the custom DMMs started with a binary buddy that concentrated more than the 90\% of the block sizes, trying to approximate the performance to the Kingsley DMM while improving the memory management in the other 10\%. It was a good policy in general, since we obtained a gain of 33.35\% in the global fitness.


\section{Conclusions and future work}
\label{sec:Conclusions}

New multimedia devices have increased their capabilities and now complex applications can be ported to them. Such applications include intensive dynamic memory requirements that must be heavily optimized for an efficient mapping on these devices. To efficiently use dynamic memory in these applications, software engineers often write custom allocators from scratch, which is a difficult and error-prone process.

In this paper, we have presented a new multi-objective optimization method based on genetic programming that can be used to optimize the complex DMMs implementations for highly dynamic applications. This method largely simplifies the exploration effort of multi-layered DMMs for designers and enables the refinement of DMM implementations in an automated way. As a result, the proposed approach leads to important savings in overall system integration time for dynamic applications. In addition, the method obtains optimal implementations of DMMs structures with respect to key designer's metrics. Moreover, our experimental results with six benchmarks and five general-purpose DMMs show that the presented optimization approach significantly reduce the execution time and memory usage consumption up to 59.27\% on average when comparing the global fitness.

The results obtained so far have outlined other interesting future research lines in the area of DMM implementation optimizations using grammatical evolution. Initially, the grammar can be extended in order to obtain much more DMM candidates. Furthermore, the study of the possible benefits of parallel algorithms in the exploration of the design space of DMM implementations is a very challenging research. Finally, other multi-objective optimization algorithms can be implemented, like optimizers with Pareto-based ranking schemes. In this work, we used a single aggregate objective function to solve this multi-objetive problem.

\appendix

\section{Profiling of the application using Pin}\label{app:PinConf}

\figurename~\ref{fig:InstrumentationCode} shows the source code that a user would write to create a Pintool that prints out a trace of address and size for every memory de/allocation in a program, as well as the time in which the function was called. To this end, we use the \texttt{RTN\_InsertCall()} function, specifying for both \texttt{malloc()} and \texttt{free()} functions the arguments of interest. The omitted code in \figurename~\ref{fig:InstrumentationCode} can be found in the Pin's \texttt{malloctrace.cpp} example.

\begin{figure}[h!]
\begin{lstlisting}[language=C++,basicstyle=\tiny\ttfamily,breaklines=true,frame=tb]
/*
  @ORIGINAL_AUTHOR: Robert Cohn
  @AUTHOR: Jos L. Risco-Martn
*/

#include "pin.H"
#include <iostream>
#include <fstream>
#include <time.h>

#if defined(TARGET_MAC)
#define MALLOC "_malloc"
#define FREE "_free"
#else
#define MALLOC "malloc"
#define FREE "free"
#endif

std::ofstream TraceFile;

KNOB<string> KnobOutputFile(KNOB_MODE_WRITEONCE, "pintool",
    "o", "MallocTrace.out", "specify trace file name");

INT32 Usage() {
    cerr << "This tool produces a trace of calls to malloc.\n";
    cerr << KNOB_BASE::StringKnobSummary();
    cerr << endl;
    return -1;
}

/* ===================================================================== */

VOID MallocBefore(CHAR * name, ADDRINT size) {
    TraceFile << name << " " << (1.0*clock())/CLOCKS_PER_SEC << " " << size;
}

VOID FreeBefore(CHAR * name, ADDRINT addr) {
    TraceFile << name << " " << (1.0*clock())/CLOCKS_PER_SEC << " " << addr << endl;
}

VOID MallocAfter(ADDRINT addr) {
    TraceFile << " " << addr << endl;
}

VOID Image(IMG img, VOID *v) {
    RTN mallocRtn = RTN_FindByName(img, MALLOC);
    if (RTN_Valid(mallocRtn)) {
        RTN_Open(mallocRtn);
        RTN_InsertCall(mallocRtn, IPOINT_BEFORE, (AFUNPTR)MallocBefore, IARG_ADDRINT, MALLOC, IARG_G_ARG0_CALLEE, IARG_END);
        RTN_InsertCall(mallocRtn, IPOINT_AFTER, (AFUNPTR)MallocAfter, IARG_G_RESULT0, IARG_END);
        RTN_Close(mallocRtn);
    }
    
    RTN freeRtn = RTN_FindByName(img, FREE);
    if (RTN_Valid(freeRtn)) {
        RTN_Open(freeRtn);
        RTN_InsertCall(freeRtn, IPOINT_BEFORE, (AFUNPTR)FreeBefore, IARG_ADDRINT, FREE, IARG_G_ARG0_CALLEE, IARG_END);
        RTN_Close(freeRtn);
    }
}

/* ===================================================================== */

VOID Fini(INT32 code, VOID *v) {
    TraceFile.close();
}

int main(int argc, char *argv[]) {
    PIN_InitSymbols();
    if( PIN_Init(argc,argv) )
        return Usage(); 

    TraceFile.open(KnobOutputFile.Value().c_str());

    IMG_AddInstrumentFunction(Image, 0);
    PIN_AddFiniFunction(Fini, 0);

    PIN_StartProgram();
    
    return 0;
}
\end{lstlisting}
\caption{DmmProfile.cpp: Instrumentation code to trace malloc() and free() function calls.}
\label{fig:InstrumentationCode}
\end{figure}

As a result, after running the application, a profiling report is available and the system designer can test different DMMs using the same profiling report. Thus, the application must be executed just once during the whole study.

\figurename~\ref{fig:InstrumentationExec} shows how the profiling of the application is performed. After including DmmProfile.cpp in the corresponding makefile, the first line in \figurename~\ref{fig:InstrumentationExec} compiles the instrumentation code. The second line runs the original target application, specified in the $<$\texttt{app}$>$ parameter, which is the original binary of the target application. Note that using this instrumentation tool, the original application does not need to be modified due to the use of dynamic libraries.

\begin{figure}[ht]
\begin{lstlisting}[basicstyle=\tiny\bf\ttfamily,frame=tb]
make dir DmmProfile.so.test

../../../pin -t obj-intel32/DmmProfile.so -- <app>
\end{lstlisting}
\caption{Profiling of the application.}
\label{fig:InstrumentationExec}
\end{figure}

\section{DMM Simulator configuration}\label{app:SimConf}

\figurename~\ref{fig:SimulatorApi} shows how to apply our library to compose a complex DMM. In this example, we first read the profiling report (lines 1-2), previously generated with Pin. When we build the \textit{ExactSegregatedFit} allocator, we provide the constructor with the minimum block size in bytes, the maximum block size in bytes and the different sizes supported by this allocator. In the example of \figurename~\ref{fig:SimulatorApi}, line 3, the last two parameters are given by the profiling report, but they can be set manually. After that, we configure the allocator defining the data structure to be used (singly-linked list), the allocation mechanism (first-fit) and the allocation policy (first in, first out). Finally, we build the corresponding DMM. As defined before, a DMM may contain one or more allocators in our case.

Finally, the simulator is invoked (lines 8-10 in \figurename~\ref{fig:SimulatorApi}) and, after a few seconds, we obtain all the metrics needed to evaluate the current DMM.

\begin{figure}[ht]
\begin{lstlisting}[language=Java,basicstyle=\tiny\ttfamily,breaklines=true,numbers=left,numberstyle=\tiny,frame=tb]
ProfilingReport profReport = new ProfilingReport();
profReport.load("profile.mem");
ExactSegregatedFit exact = new ExactSegregatedFit(0, profReport.getMaxSizeInB(), profReport.getSizesInB());
exact.setup(FreeList.DATA_STRUCTURE.SLL, FreeList.ALLOCATION_MECHANISM.FIRST, FreeList.ALLOCATION_POLICY.FIFO);

DynamicMemoryManager manager = new DynamicMemoryManager(exact);

Simulator simulator = new Simulator(profReport, manager);
simulator.start();
simulator.join();
\end{lstlisting}
\caption{Configuration of a DMM using the Simulator API}
\label{fig:SimulatorApi}
\end{figure}

At the same time the simulation runs, several relevant metrics are computed, such as number of de/allocations, splittings, coalescings, performance, memory usage, memory accesses, etc. All the previous parameters except the execution time can be calculated accurately. However, since the system is using simulation time instead of real time, the total execution time is calculated as the computational complexity or time complexity \cite{Sipser1996}.

\begin{figure}[t!]
\begin{lstlisting}[language=Java,basicstyle=\tiny\ttfamily,breaklines=true,frame=tb]
void firstFit(long sizeInB) {
 // ...
 while(iterator.hasNext()) {
  counterForMetrics++;
  currentBlock = iterator.next();
  if(currentBlock.sizeInB>=sizeInB) {
   block = currentBlock;
   iterator.remove();
   break;
  }
 }
 metrics.addExecutionTime(counterForMetrics);
 metrics.addMemoryAccesses(2*counterForMetrics);
 // ...
}
\end{lstlisting}
\caption{Execution time and memory accesses computation}
\label{fig:ExTimeComputation}
\end{figure}

The code snippet in \figurename~\ref{fig:ExTimeComputation} shows an illustrative example of how this task is performed in the proposed DMM optimization framework. This code excerpt shows a portion of the first-fit algorithm inside the simulator. The main loop looks for the first block big enough to allocate the requested size. We count the number of iterations in the loop, and after that, both the execution time and memory accesses are updated accordingly (+1 for each cycle in the loop to consider the computational time and +2 for each cycle to count two accesses in the actual allocator: one to the current node in the free-list and another one to compute the size, i.e., subtraction of two pointers).

\begin{figure}[t!]
\centering
\includegraphics[width=0.55\columnwidth]{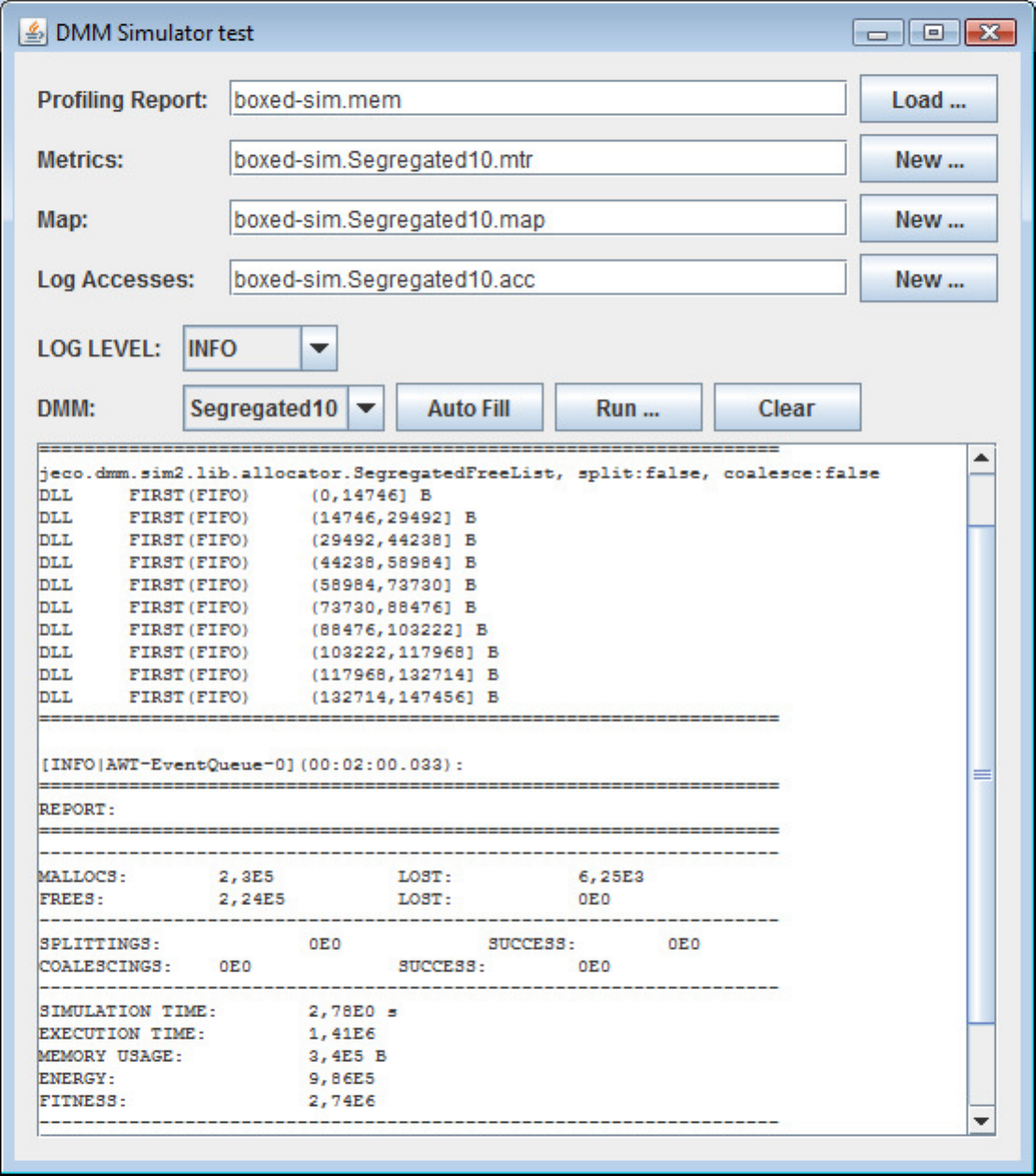}
\caption{Simulator Graphical User Interface}
\label{fig:SimulatorGui}
\end{figure}

Finally, as \figurename~\ref{fig:SimulatorGui} shows, to facilitate the use of the simulator, we have developed a GUI to test some general-purpose memory allocators, as well as to perform an automatic exploration of DMMs for the target application. Given a profiling report, the interface simulates the selected allocator, giving its ``map'' and some of the computed metrics. In any case, all the metrics are saved in external files.


\section*{References}
\bibliographystyle{elsarticle-num}
\bibliography{bibliography}

\begin{thebibliography}{10}
\expandafter\ifx\csname url\endcsname\relax
  \def\url#1{\texttt{#1}}\fi
\expandafter\ifx\csname urlprefix\endcsname\relax\def\urlprefix{URL }\fi
\expandafter\ifx\csname href\endcsname\relax
  \def\href#1#2{#2} \def\path#1{#1}\fi

\bibitem{Johnstone1999}
M.~S. Johnstone, P.~R. Wilson, The memory fragmentation problem: solved?,
  SIGPLAN Not. 34~(3) (1999) 26--36.
\newblock \href {https://doi.org/http://doi.acm.org/10.1145/301589.286864}
  {\path{doi:http://doi.acm.org/10.1145/301589.286864}}.

\bibitem{Lea}
Doug {L}ea. {A} memory allocator., \\http://g.oswego.edu/dl/html/malloc.html
  (2010).

\bibitem{Barrett1993}
D.~A. Barrett, B.~G. Zorn, Using lifetime predictors to improve memory
  allocation performance, SIGPLAN Not. 28~(6) (1993) 187--196.
\newblock \href {https://doi.org/http://doi.acm.org/10.1145/173262.155108}
  {\path{doi:http://doi.acm.org/10.1145/173262.155108}}.

\bibitem{Grunwald1993}
D.~Grunwald, B.~Zorn, Custo{M}alloc: efficient synthesized memory allocators,
  Softw. Pract. Exper. 23~(8) (1993) 851--869.

\bibitem{SPEC2010}
SPEC, Standard {P}erformance {E}valuation {C}orporation, Available at:
  http://www.spec.org (2013).

\bibitem{Berger2001}
E.~D. Berger, B.~G. Zorn, K.~S. McKinley, Composing high-performance memory
  allocators, in: PLDI '01: Proceedings of the ACM SIGPLAN 2001 conference on
  Programming language design and implementation, ACM, New York, NY, USA, 2001,
  pp. 114--124.
\newblock \href {https://doi.org/http://doi.acm.org/10.1145/378795.378821}
  {\path{doi:http://doi.acm.org/10.1145/378795.378821}}.

\bibitem{Calder1995}
B.~Calder, D.~Grunwald, B.~Zorn, Quantifying behavioral differences between {C}
  and {C}++ programs, Journal of Programming Languages 2 (1995) 313--351.

\bibitem{Atienza2006a}
D.~Atienza, J.~M. Mendias, S.~Mamagkakis, D.~Soudris, F.~Catthoor, Systematic
  dynamic memory management design methodology for reduced memory footprint,
  ACM Trans. Des. Autom. Electron. Syst. 11~(2) (2006) 465--489.
\newblock \href {https://doi.org/http://doi.acm.org/10.1145/1142155.1142165}
  {\path{doi:http://doi.acm.org/10.1145/1142155.1142165}}.

\bibitem{Vijaykrishnan2003}
N.~Vijaykrishnan, M.~Kandemir, M.~J. Irwin, H.~S. Kim, W.~Ye, D.~Duarte,
  Evaluating integrated hardware-software optimizations using a unified energy
  estimation framework, IEEE Trans. Comput. 52~(1) (2003) 59--76.
\newblock \href {https://doi.org/http://dx.doi.org/10.1109/TC.2003.1159754}
  {\path{doi:http://dx.doi.org/10.1109/TC.2003.1159754}}.

\bibitem{Diaz2011}
J.~Díaz, J.~M. Colmenar, J.~L. Risco-Mart\'{i}n, J.~L. Ayala, O.~Garnica,
  {Quantifying the Impact of Dynamic Memory Managers into Memory-Intensive
  Applications}, in: SCSC 2011: Summer Computer Simulation Conference, 2011,
  pp. 1--10.

\bibitem{RiscoMartin2009b}
J.~L. Risco-Martin, D.~Atienza, R.~Gonzalo, J.~I. Hidalgo, Optimization of
  dynamic memory managers for embedded systems using grammatical evolution, in:
  GECCO '09: Proceedings of the 11th Annual conference on Genetic and
  evolutionary computation, ACM, New York, NY, USA, 2009, pp. 1609--1616.
\newblock \href {https://doi.org/http://doi.acm.org/10.1145/1569901.1570116}
  {\path{doi:http://doi.acm.org/10.1145/1569901.1570116}}.

\bibitem{Meyers1995}
S.~Meyers, More effective {C}++: 35 new ways to improve your programs and
  designs, Addison-Wesley Longman Publishing Co., Inc., 1995.

\bibitem{Atienza2006}
D.~Atienza, S.~Mamagkakis, F.~Poletti, J.~M. Mendias, F.~Catthoor, L.~Benini,
  D.~Soudris, Efficient system-level prototyping of power-aware dynamic memory
  managers for embedded systems, Integr. VLSI J. 39~(2) (2006) 113--130.
\newblock \href {https://doi.org/http://dx.doi.org/10.1016/j.vlsi.2004.08.003}
  {\path{doi:http://dx.doi.org/10.1016/j.vlsi.2004.08.003}}.

\bibitem{Lo2004}
C.-T.~D. Lo, W.~Srisa-an, J.~M. Chang,
  \href{http://dx.doi.org/10.1016/S0164-1212(03)00095-5}{The design and
  analysis of a quantitative simulator for dynamic memory management}, J. Syst.
  Softw. 72 (2004) 443--453.
\newblock \href
  {https://doi.org/http://dx.doi.org/10.1016/S0164-1212(03)00095-5}
  {\path{doi:http://dx.doi.org/10.1016/S0164-1212(03)00095-5}}.
\newline\urlprefix\url{http://dx.doi.org/10.1016/S0164-1212(03)00095-5}

\bibitem{Teng2008}
G.~Teng, K.~Zheng, W.~Dong, Sdma: A simulation-driven dynamic memory allocator
  for wireless sensor networks, in: Sensor Technologies and Applications, 2008.
  SENSORCOMM '08. Second International Conference on, 2008, pp. 462--467.
\newblock \href {https://doi.org/10.1109/SENSORCOMM.2008.9}
  {\path{doi:10.1109/SENSORCOMM.2008.9}}.

\bibitem{RiscoMartin2010a}
J.~L. Risco-Martín, J.~M. Colmenar, D.~Atienza, J.~I. Hidalgo, Simulation of
  high-performance memory allocators, in: Proccedings of the 13th EUROMICRO
  Conference on Digital System Design, 2010, pp. 275--282.
\newblock \href {https://doi.org/10.1109/DSD.2010.44}
  {\path{doi:10.1109/DSD.2010.44}}.

\bibitem{RiscoMartin2011a}
J.~L. Risco-Martín, J.~M. Colmenar, D.~Atienza, J.~I. Hidalgo,
  \href{http://linkinghub.elsevier.com/retrieve/pii/S0141933111000937}{Simulation
  of high-performance memory allocators}, Microprocessors and Microsystems
  35~(8) (2011) 755--765.
\newblock \href {https://doi.org/10.1016/j.micpro.2011.08.003}
  {\path{doi:10.1016/j.micpro.2011.08.003}}.
\newline\urlprefix\url{http://linkinghub.elsevier.com/retrieve/pii/S0141933111000937}

\bibitem{RiscoMartin2010b}
J.~L. Risco-Martín, D.~Atienza, J.~M. Colmenar, O.~Garnica, A parallel
  evolutionary algorithm to optimize dynamic memory managers in embedded
  systems, Parallel Computing 36~(10) (2010) 572--590.
\newblock \href {https://doi.org/10.1016/j.parco.2010.07.001}
  {\path{doi:10.1016/j.parco.2010.07.001}}.

\bibitem{Jones2012}
R.~Jones, A.~Hosking, E.~Moss, The Garbage Collection Handbook, Chapman \&
  Hall, 2012.

\bibitem{Wilson1995}
P.~R. Wilson, M.~S. Johnstone, M.~Neely, D.~Boles, Dynamic storage allocation:
  A survey and critical review, in: IWMM '95: Proceedings of the International
  Workshop on Memory Management, Springer-Verlag, London, UK, 1995, pp. 1--116.

\bibitem{Berger2002}
E.~D. Berger, B.~G. Zorn, K.~S. McKinley, Reconsidering custom memory
  allocation, in: Proceedings of the 17th ACM SIGPLAN conference on
  Object-oriented programming, systems, languages, and applications, OOPSLA
  '02, ACM, New York, NY, USA, 2002, pp. 1--12.
\newblock \href {https://doi.org/10.1145/582419.582421}
  {\path{doi:10.1145/582419.582421}}.

\bibitem{PABA}
J.~L. Risco-Martín, J.~M. Colmenar, Parallel {A}rchitectures and {B}ioinspired
  {A}lgorithms ({PABA}) research group - {S}ource code repository, Available
  at: https://code.google.com/p/paba/ (2008).

\bibitem{Luk2005}
C.-K. Luk, R.~Cohn, R.~Muth, H.~Patil, A.~Klauser, G.~Lowney, S.~Wallace, V.~J.
  Reddi, K.~Hazelwood, Pin: building customized program analysis tools with
  dynamic instrumentation, in: PLDI '05: Proceedings of the 2005 ACM SIGPLAN
  conference on Programming language design and implementation, ACM, New York,
  NY, USA, 2005, pp. 190--200.
\newblock \href {https://doi.org/http://doi.acm.org/10.1145/1065010.1065034}
  {\path{doi:http://doi.acm.org/10.1145/1065010.1065034}}.

\bibitem{ONeill2003}
M.~O'Neill, C.~Ryan, Grammatical Evolution: Evolutionary Automatic Programming
  in an Arbitrary Language, Kluwer Academic Publishers, 2003.

\bibitem{Brabazon2006}
A.~Brabazon, M.~O'Neill, Biologically Inspired Algorithms for Financial
  Modelling, Springer, 2006.

\bibitem{Dempsey2007}
I.~Dempsey, M.~O'Neill, A.~Brabazon, Constant creation in grammatical
  evolution, Int. J. Innov. Comput. Appl. 1~(1) (2007) 23--38.
\newblock \href {https://doi.org/http://dx.doi.org/10.1504/IJICA.2007.013399}
  {\path{doi:http://dx.doi.org/10.1504/IJICA.2007.013399}}.

\bibitem{ONeill2001}
M.~O'Neill, C.~Ryan, Grammatical evolution, IEEE Trans. Evolutionary
  Computation 5~(4) (2001) 349--358.

\bibitem{Brabazon2008}
A.~Brabazon, M.~O'Neill, I.~Dempsey, An introduction to evolutionary
  computation in finance, Computational Intelligence Magazine, IEEE 3~(4)
  (2008) 42--55.
\newblock \href {https://doi.org/10.1109/MCI.2008.929841}
  {\path{doi:10.1109/MCI.2008.929841}}.

\bibitem{Poli2008}
R.~Poli, W.~B. Langdon, N.~F. McPhee, A field guide to genetic programming,
  Published via http://lulu.com and freely available at
  http://www.gp-field-guide.org.uk., 2008.

\bibitem{SPECpu2006}
SPEC, Standard {P}erformance {E}valuation {C}orporation - cpu2006, Available
  at: http://www.spec.org/cpu2006 (2013).

\bibitem{Sipser1996}
M.~Sipser, Introduction to the Theory of Computation, International Thomson
  Publishing, 1996.

\end{thebibliography}

\end{document}